\documentclass[aps,prl,reprint,superscriptaddress]{revtex4-1}
\usepackage{graphicx}
\usepackage{amsmath}
\usepackage{multirow}
\begin{document}


\title{Exciting Reflectionless Unidirectional Edge Modes in a Reciprocal Photonic Topological Insulator Medium}


\author{Bo Xiao}
\affiliation{Department of Electrical and Computer Engineering, University of Maryland, College Park, Maryland 20742-3285, USA}
\affiliation{Center for Nanophysics and Advanced Material, Physics Department, University of Maryland, College Park, Maryland 20742-3285, USA}
\author{Kueifu Lai}
\affiliation{Physics Department, University of Texas at Austin, Austin, Texas 78712, USA}
\author{Yang Yu}
\affiliation{Physics Department, University of Texas at Austin, Austin, Texas 78712, USA}
\author{Tzuhsuan Ma}
\affiliation{Physics Department, University of Texas at Austin, Austin, Texas 78712, USA}
\author{Gennady Shvets}
\affiliation{Physics Department, University of Texas at Austin, Austin, Texas 78712, USA}
\author{Steven M. Anlage}
\affiliation{Department of Electrical and Computer Engineering, University of Maryland, College Park, Maryland 20742-3285, USA}
\affiliation{Center for Nanophysics and Advanced Material, Physics Department, University of Maryland, College Park, Maryland 20742-3285, USA}

\date{\today}

\begin{abstract}
Photonic topological insulators are an interesting class of materials whose photonic band structure can have a bandgap in the bulk while supporting topologically protected unidirectional edge modes. Recent studies \cite{PhysRevLett.114.127401, PhysRevLett.114.037402, Chen2014, kueifu2016, Khanikaev2013, Cheng2016} on bianisotropic metamaterials that emulate the electronic quantum spin Hall effect using its electromagnetic analog are examples of such systems with relatively simple and elegant design. In this paper, we present a rotating magnetic dipole antenna, composed of two perpendicularly oriented coils, that can efficiently excite the unidirectional topologically protected surface waves in the bianisotropic metawaveguide (BMW) structure recently realized by Ma, et al.  \cite{PhysRevLett.114.127401}, despite the fact that the BMW medium does not break time-reversal invariance. In addition to achieving high directivity, the antenna can be tuned continuously to excite reflectionless edge modes to the two opposite directions with various amplitude ratios. We demonstrate its performance through experiment and compare to simulation results.
\end{abstract}


\maketitle

Topological insulators \cite{RevModPhys.82.3045,RevModPhys.83.1057} are a class of materials that are insulating in the bulk and are conducting only on the edge or surface. These materials have attracted much research effort because of the robust transport properties of the edge states in the presence of impurities and disorder in the material. One example of such a system is the integer quantum Hall effect (QHE) in two-dimensional semiconductors, discovered by von Klitzing in 1980 \cite{PhysRevLett.45.494}, where edge states are unidirectional and reflectionless. The electronic QHE systems require very low temperatures and strong magnetic fields and thus are difficult to realize. However its electromagnetic counterpart, photonic systems that emulate the spin of electrons and the effects of magnetic field or spin-orbit interaction, have recently enjoyed an explosion of interest \cite{PhysRevLett.100.013904, PhysRevA.78.033834, Hafezi2011, Fang2012, Khanikaev2013, Lu2013, PhysRevLett.113.113904, PhysRevLett.109.106402, Rechtsman2013, HafeziM.2013, Lu2014}. Three main types of electromagnetic analog systems have been presented so far: magnetic photonic crystals \cite{PhysRevLett.100.013905, Wang2009}, coupled resonators and waveguides \cite{Rechtsman2013, Hafezi2011, HafeziM.2013, PhysRevLett.113.087403} and bianisotropic metamaterials \cite{PhysRevLett.114.127401, PhysRevLett.114.037402, Chen2014, kueifu2016, Khanikaev2013, Cheng2016}. In this paper, we are interested in exciting and measuring the topologically protected surface waves (TPSWs) in the bianisotropic metawaveguide (BMW) structure recently realized by Ma, et al. \cite{PhysRevLett.114.127401}. This metawaveguide supports photonic modes that have the same topological nature as the electronic states in graphene with strong spin-orbit coupling, as described by the Kane-Mele Hamiltonian \cite{PhysRevLett.95.226801, PhysRevLett.96.106802}. A more distinguishing feature of this BMW is that it does not require broken time-reversal symmetry, as utilized in previous studies \cite{PhysRevLett.106.093903,Wang2009,PhysRevLett.100.013904}, thus bringing a new genre of topological insulators. Furthermore, the BMW structure is based on the quantum spin-Hall effect, and can be scaled to higher frequencies, far beyond the regime where ferrite-based non-reciprocal systems will work. Here we demonstrate for the first time excitation of a uni-directional TPSW in this reciprocal medium.

The design of the quantum spin Hall analog BMW structure follows three steps \cite{Khanikaev2013, PhysRevLett.114.037402}. It begins with a parallel plate waveguide filled with metal rods regularly arranged in a hexagonal graphene-like lattice that connect the upper and lower plates. The dimensions, such as the rod radius, height and lattice spacing, are carefully tuned such that the transverse electric (TE) and magnetic (TM) propagating electromagnetic modes are degenerate at the K (K') point in the Brillouin zone, where there is a Dirac point in the photonic band structure. This degeneracy is essential to creating a spin-like degree of freedom, which can be interpreted as the phase relationship between TE and TM modes of the metamaterial, in-phase for the spin-up and out-of-phase for the spin-down states.   In the second step, a symmetry-breaking air gap between the metal rods and the top plate is introduced, creating bi-anisotropy and forming a bandgap at the K(K') point that provides the photonic insulating behavior in the bulk. Finally, combining two such BMW structures, one with an air gap on the top plate and another one with an air gap on the bottom plate, forms an interface that supports the TPSWs. The TPSWs exhibit reflectionless unidirectional propagation in first-principles simulation using COMSOL, and also in a recent experiment \cite{kueifu2016}. In this paper, we experimentally demonstrate the launching of unidirectional TPSWs on such an interface by means of a rotating magnetic dipole antenna. It was suggested in \cite{PhysRevLett.114.127401} that the BMW is also chiral, i.e. it enables unidirectional excitation of TPSWs by a circularly polarized electric or magnetic dipole. While other chiral photonic waveguides have been recently realized \cite{Sollner2015}, the proposed BMW would represent the first example of a chiral photonic circuit with topological protection against back-reflections without the use of ferrite or synthetic gauge fields to break time-reversal symmetry.  

\begin{figure}
\includegraphics[width=0.4\textwidth]{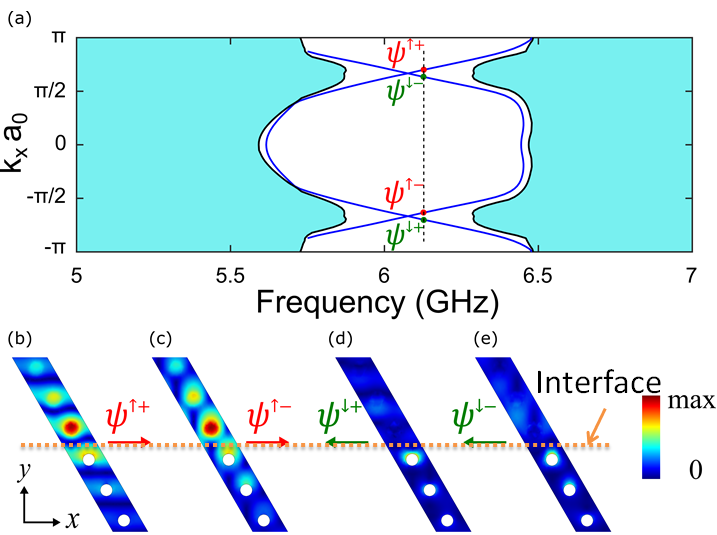}
\caption{\label{fig:bo}(a): Numerically calculated band structure of the waveguide, where four edge modes have been labeled. (b) to (e): Field profile $|H_x-iH_y|$ of the four edge modes at 6.13GHz (shown in (a) as a dashed line). It is clear that only the two spin-up states (from the two valleys), but none of the spin-down states, can be excited by a right circularly polarized magnetic dipole.}
\end{figure}

A BMW with an interface between two topologically nontrivial domains is constructed as in Ref. \cite{PhysRevLett.114.127401, kueifu2016} with waveguide height $h_0=36.8$mm, rod diameter $d_0=12.7$mm, period $a_0=36.8$mm, air gap size $g_0=5.5$mm. The designed center frequency for the bulk insulating bandgap is around 6.08 GHz, as shown in Fig. \ref{fig:bo} (a). It has been shown in simulation that TPSWs propagate along the interface with high transmission ($T>0.9$) over nearly the entire bulk bandgap \cite{PhysRevLett.114.127401}, and also in experiment that the TPSWs are observed to boost the transmission by nearly 30 dB between 5.87 GHz and 6.29 GHz compared to the bulk transmission \cite{kueifu2016}. The source antenna used in \cite{kueifu2016} is a linear dipole antenna which excites both left-going and right-going TPSWs. Our goal is to excite TPSWs that only propagate towards one direction and to further control the directivity and relative amplitude of the excited waves, and also to demonstrate that a circularly polarized (CP) dipole placed inside an airgap of the BMW structure excites a unidirectional guided wave.
\par In order to efficiently excite the TPSW in the BMW structure, the source must be placed at a location where the edge mode field is most intense and it must generate a field profile that matches with the edge mode at that location. By inspecting the numerically calculated intensity profiles of the magnetic field projected onto the circularly polarized basis in Fig. \ref{fig:bo} (b) - (e) for the spin-up and spin-down eigenmodes belonging to the two valleys of the Brillouin zone, the 5.5mm air gap between the metal plate and cylinder rod was identified as the best place to locate the source. At that location the field profile is a rotating magnetic dipole around the center frequency of 6.08 GHz. The right circularly polarized (RCP) component of the magnetic field is maximized for the forward-propagating spin-up eigenmodes, and vanishes for the backward-propagating spin-down eigenmodes. This implies chirality, i.e that an RCP magnetic dipole should excite only forward TPSWs. The spin-up and spin-down surface modes have opposite handedness for the rotating magnetic field at this location \cite{PhysRevLett.114.127401}. As shown in Fig. \ref{fig:bo}, a RCP magnetic dipole only excites the right-going modes. The direction of propagation is locked to the spin state, allowing for uni-directional TPSW propagation depending on the sense of rotation of the magnetic dipole antenna. Mechanically rotating a loop antenna at a rate on the order of $10^9$ revolutions per second is too difficult to achieve. Instead, we use two loop antennas, perpendicularly positioned to each other, with two $\sim6$ GHz sinusoidal input voltage waves with variable phase shift as $v_A(t)=V_A\cos(\omega t)$, $v_B(t)=V_B\cos(\omega t + \phi)$ to antennas A and B, respectively.  By setting $V_A=V_B$ and $\phi=+(-) \pi/2$, a clockwise (or counter-clockwise) rotating magnetic dipole is formed. Furthermore, by tuning $\phi$ one can continuously vary the amplitude of the left or right going modes, providing more control over the directivity.

\par Circularly polarized (CP) magnetic dipole antennas have been introduced in several fields. In plasma physics, magnetic dipole antennas are used to study the interaction of rotating magnetic fields with plasma \cite{Karavaev3274916,rmfsource}. And in magnetic resonance imaging technology, two mutually perpendicular RF coils, also called quadrature coils, can improve sensitivity by up to 41\% and reduce power consumption by half compared to linear RF coils \cite{HAYES1985622, GLOVER1985255}. In our case two perpendicularly placed RF coils is quite effective and simple to implement. The arrangement of this antenna is similar to the turnstile antennas (also known as crossed dipole antennas) \cite{h1937antenna, griffee1974antenna, kraus1988antennas}, widely used in satellite communications, consisting of two crossed electric dipoles fed with quadrature phase shift. However the electric dipole in the turnstile and the magnetic dipole in our design are very different in nature.

\begin{figure}
\includegraphics[width=0.4\textwidth]{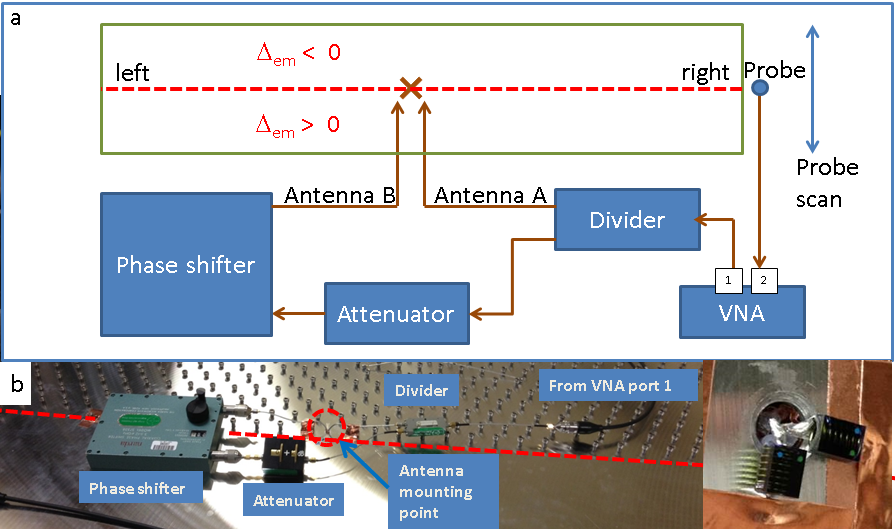}
\caption{\label{fig:schema}(a) Schematic of the experimental setup. 
$\Delta_{em}$ is the bianisotropy coefficient \cite{PhysRevLett.114.127401} and the interface is defined by the boundary of the two regions with $\Delta_{em}$ of opposite sign, denoted by a red dashed line. (b) Photograph of the top of the BMW structure showing the corresponding elements in (a).
The inset shows the arrangement of the two loop coils 
which are in the air gap of a cylinder at the interface.}
\end{figure}

A rectangular BMW structure is created (1.7 meters by 0.74 meters), consisting of 45 by 20 unit cells, with a single interface bisecting the structure in the length-wise direction, as shown in Figure \ref{fig:schema}.  The rotating dipole antenna is placed inside the BMW structure near the middle of the interface.  The source of microwave signals is the first port of an Agilent E5071C Vector Network Analyzer (VNA), and that signal is split approximately in half by a HP87304C power divider, creating two branches. One branch goes directly to loop antenna A while the other one is connected to a variable attenuator and variable phase shifter before going to the other loop antenna, B. The two loop antennas are in fact air-core RF coil inductors (Coilcraft model number 1812SMS-56NJLB, $R$=6.2 mOhm, $L$=56 nH at 150 MHz) with one end grounded to the metal plate and the other end soldered onto the center conductor of the feed coaxial cable, as shown in the inset of Figure \ref{fig:schema}. Hence the input signal from the VNA is divided, phase-shifted and then sent to the two RF coils that each creates a magnetic dipole. By controlling the value of the phase shift, it can produce a linearly, circularly or elliptically polarized magnetic dipole source.

\par
\begin{figure}
\includegraphics[width=0.5\textwidth]{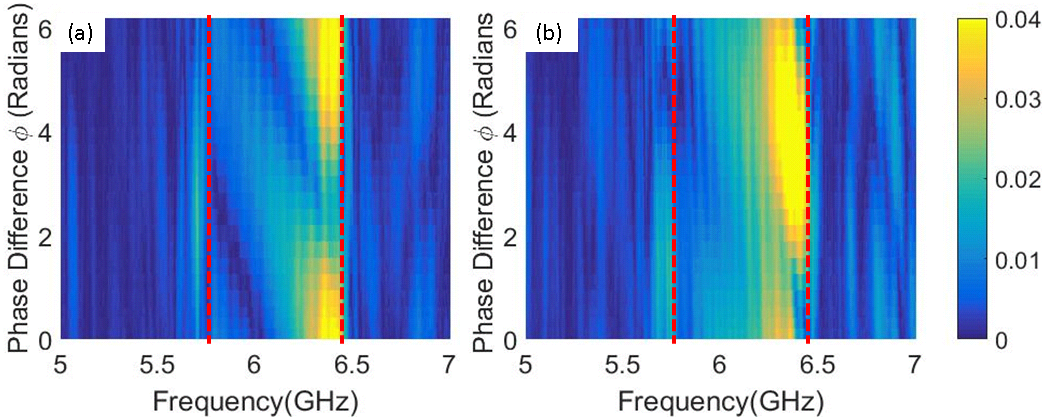}
\caption{\label{fig:fig1}Transmission amplitude at the (a) left and (b) right side of the BMW interface as a function of frequency while varying the phase difference $\phi$ of the two driving loop antennas
. The probe is positioned at the center of the edge. The BMW bulk band gap extends from 5.80 GHz to 6.47 GHz, as shown with the vertical dashed lines. }
\end{figure}

 A transmission experiment is performed by placing a simple electric dipole antenna \cite{PhysRevE.74.036213} at the edge of the BMW structure on either the left or right side where the interface comes to the edge to pick up and record the transmitted signal at port 2 of the VNA. We then move the probe along the edge to do a lateral scan (see Fig. \ref{fig:schema}(a)), recording the transmission amplitude as a function of the probe's location and also the phase difference between the two loops. The experimental result for the transmission when the probe is at the midpoint of the edge is shown in Fig \ref{fig:fig1}. The transmission data when we vary both the probe location and the phase difference can be found in Supplementry Material Movie 1. From the midpoint transmission plot, the excited wave is propagating primarily to the left (right) when the phase difference is close to $2\pi$ ($1.27\pi$) confirming the successful excitation of a predominantly unidirectional edge mode. It is also clear that the transmission has a period of $2\pi$ with respect to the phase difference in the frequency range of 5.80 GHz to 6.47 GHz, which corresponds to the plateau of enhanced transmission due to TPSWs in Figure 1 of Ref. \cite{kueifu2016}. The TPSWs are more efficiently excited at higher frequencies, with a peak at 6.47 GHz. The Supplementry Material Movie 1 further demonstrates that the TPSWs are propagating along the interface, resulting in a spatial focus around the center of the edge, and that the left- and right-going TPSWs are controlled by varying the phase difference $\phi$.

Naively, it is expected that when $V_A=V_B$ and $\phi=\pi/2$, a clockwise rotating dipole would excite a purely left or right going edge mode, depending on whether the dipole is place at the bottom or upper air gap, and thus the left transmission should reach its maximum while the right transmission should be zero. However, the experiment is affected by several non-idealities beyond our control. The two loop antennas are different in their geometry, the coupling to the edge mode, and the actual power received from input feed lines. Hence we must loosen the restrictions on $V_A$, $V_B$ and $\phi$ to obtain a more general result. To keep the total input power constant while giving the two loop antenna variable portions of the power, we parameterize the amplitudes of the driving voltage waves with an angle $\theta$ as $V_A=V_0\cos\theta$ and $V_B=V_0\sin\theta$, where $\theta\in[0,\pi/2]$. 

We performed a numerical simulation of a finite-size BMW structure, whose detailed geometry can be found in the Supplementry Material. The simulation calculates the transmission from antenna A and B to the left and right side probes in Fig. \ref{fig:schema} and expresses it as a 2-by-2 scattering matrix $\mathbf{S}_{\textnormal{sim}}=\begin{bmatrix}
 S_{LA} & S_{LB}\\
S_{RA} & S_{RB}
\end{bmatrix}$, where $S_{ij}$ is the transmission from $j$ to $i$, $L (R)$ represent left (right) side probe, $A (B)$ represent antenna A (B) respectively.

\begin{figure}
\includegraphics[width=0.4\textwidth]{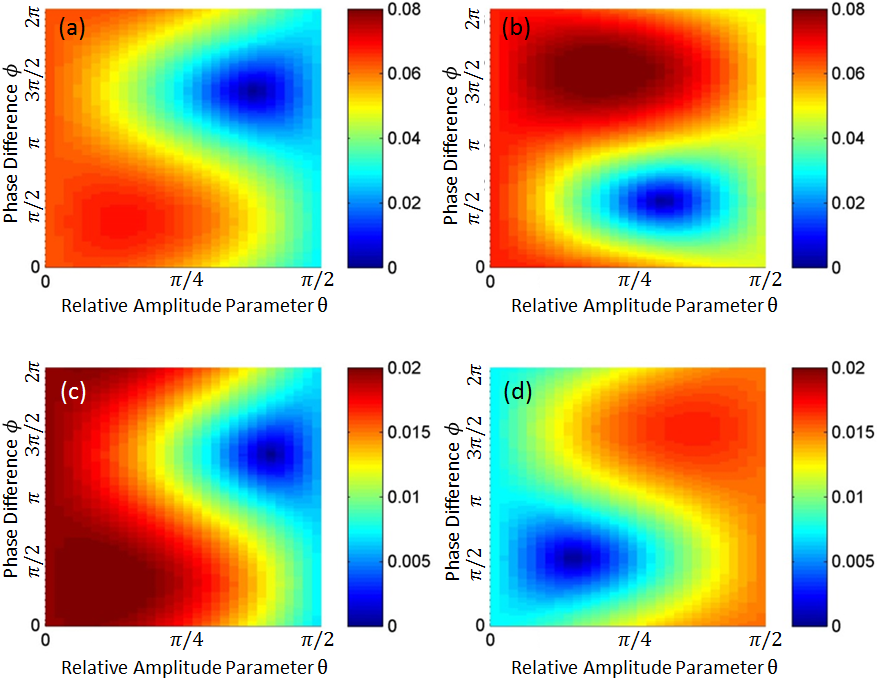}
\caption{\label{fig:fig2}CST simulation results for transmission amplitude to the left and right side of the BMW structure while varying both the phase difference $\phi$ and driving amplitude (parameterized by angle $\theta \in [0, \pi/2]$) of the two loop antennas.  Results are shown for (a) left, (b) right 6.47 GHz and (c) left, (d) right 6.08 GHz.}
\end{figure}

Given the simulation result of $\mathbf{S}_{\textnormal{sim}}$, and that $V_A=V_0\cos\theta$, $V_B=V_0\sin\theta$ (to keep the total input power constant), we can calculate the transmission to the left or right side with different driving amplitudes in the two antennas ($\theta$) and different phase shift values ($\phi$). To calculate transmission to the right side for instance one has $V_R=V_A S_{RA} + V_Be^{-i\phi} S_{RB} =V_0(S_{RA}\cos \theta+e^{-i\phi}S_{RB}\sin \theta)$ where $e^{-i\phi}$ controls the phase difference between the A, B antenna. Figure \ref{fig:fig2} shows the resulting transmission to the left and right probes as a function of $\theta$ and $\phi$, at two different frequencies within the bulk bandgap of the BMW structure. Since $\mathbf{S}$ is frequency dependent and is simulated over the whole bandgap, choosing a different frequency could lead to a different plot but the unidirectional propagation property should remain. Focusing on the minimum and maximum values of the transmission, we observe a number of key features:
\begin{itemize}
\item The transmission has a period of $2\pi$ with phase difference $\phi$.
\item When the transmission to the left side is at its maximum, the transmission to the right side is not zero and is not exactly at its minimum (but is near it).
\item If $\theta$ and $\phi$ are chosen wisely, one can effectively eliminate the transmission to one side although the transmission to the other side is not at its maximum.
\end{itemize}

\begin{figure}
\includegraphics[width=0.4\textwidth]{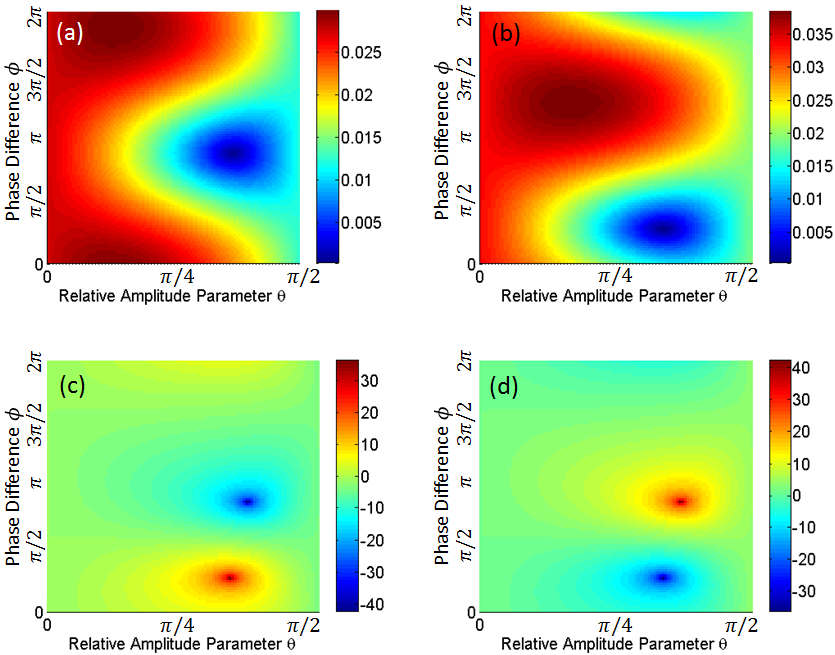}
\caption{\label{fig:fig3}Transmission amplitude to the (a) left and (b) right side when varying both the phase difference $\phi$ and the input power (parameterized by $\theta$) of the two loop antennas as deduced from the data at a frequency of 6.38 GHz.  To examine the directivity, we plot the ratio of (c) left to right side and (d) right to left side transmission amplitude (in dB) as a function of $\theta$ and $\phi$.}
\end{figure}

Note that all these observations regarding the results of the simulations are in agreement with measurements. To more firmly connect the simulation results with the experimental data, we followed these steps:
\begin{enumerate} 
\item Model the transmission process in the experiment using a 2-by-2 scattering matrix, $\mathbf{S}_{\textnormal{exp}}=
\begin{bmatrix}
S_{LA} & S_{LB} \\
S_{RA} & S_{RB}
\end{bmatrix}
$. The experimental data can then be expressed as
\begin{equation}
\label{equ:exp}
\begin{bmatrix}
L_1 & L_2 & \hdots & L_N \\
R_1 & R_2 & \hdots & R_N
\end{bmatrix} = 
\frac{1}{\sqrt{2}}
\mathbf{S}_{\textnormal{exp}}
\begin{bmatrix}
1 & 1 & \hdots & 1 \\
e^{-i\phi_1} & e^{-i\phi_2} & \hdots & e^{-i\phi_N}  
\end{bmatrix},
\end{equation}
where $\phi_i$ ($i=1,2,...,N$) are the $N=31$ known experimental values of the loop antenna phase differences, and $L_i$ $R_i$ are the measured complex transmission data taken at the left and right sides of the BMW interface edge.  It is assumed that the amplitudes of the driving voltage on the two antennas are both $V_0/\sqrt{2}$.  All variables in Eq. \ref{equ:exp} are functions of frequency.
\item Fit the experimental data to this model to obtain the $\mathbf{S}_{\textnormal{exp}}$ matrix as a function of frequency.  This fit is strongly over-determined, but is found to be very good for all frequencies (with normalized mean-square-error of around 7\%) and both left or right side transmission.
\item Calculate the expected left side transmission for $M=91$ values of $\theta$ and $N=31$ values of $\phi$ using $L_{nm}=S_{LA}\cos\theta_m+e^{-i\phi_n}S_{LB}\sin\theta_m$, where $n=1,2,...,N$, $m=1,2,...,M$. 

The same calculations are done for right side transmission, and the result for transmission amplitude is shown in Figure \ref{fig:fig3} (a) and (b).
\end{enumerate}

It is clear that Fig.\ref{fig:fig2} (simulation) and Fig.\ref{fig:fig3} (deduced from experimental data) have very similar patterns.  By choosing the appropriate $\theta$ and $\phi$ values, one can minimize the one side transmission amplitude to as low as 0.0002 or tune the ratio of the left to right transmission amplitude (Fig.\ref{fig:fig3} (c) and (d)) from -42 dB to +36 dB. A summary of the results for extreme values of transmission is given in Table \ref{table1}.  All of this confirms that the rotating magnetic dipole antenna is an effective way to excite directional edge modes in the BMW structure with tunable directivity.

\begin{table}
\begin{tabular}{| l | l | l | l | l | l | l | }
\hline
  & & $\phi$ (degrees) & $\theta$ (degrees)& $T_L$ & $T_R$ & $20\log(\frac{T_L}{T_R})$ \\
\hline
\multirow{2}{*}{Left} & Max & 334.5 & 23.6 & 0.0297 & 0.0293 & 0.118 dB \\
& Min & 156.4 & 65.5 & 0.0002 & 0.0254 & -42.1 dB\\
\hline
\multirow{2}{*}{Right} & Max & 229.1 & 30.0 & 0.0225 & 0.0384 & -4.64 dB \\
& Min & 47.3 & 60.0 & 0.0197 & 0.0003 & 36.3dB \\
\hline
\end{tabular}
\caption{\label{table1} Summary of extreme transmisison values as deduced from experimental data where $T_L$ and $T_R$ are transmission amplitude to left and right, respectively.}
\end{table}

In terms of applications, our results can be used for selective feeding of a waveguide in either direction.  This can be used to feed a beam-forming array of antennas through a series of sensitive and rapidly tunable structures.  The current design can also handle high microwave powers making it attractive for transmit applications. Since the directivity of the edge modes can be varied by $\phi$, it can be used as a modulation method for communications.

In conclusion we have experimentally demonstrated excitation of a unidirectional edge mode using a rotating magnetic dipole antenna consisting of two perpendicular coils. The edge mode in this time-reversal symmetry preserved Bianisotropic Meta-waveguide has been demonstrated to be unidirectional to the level of one part in $10^{4}$.  In addition, the degree of directionality can be tuned continuously using the method that we have outlined here, allowing for novel applications in the field of communications, for example phased array antennas.

\begin{acknowledgments}
This work was supported by ONR under Grant No. N000141512134, AFOSR COE Grant FA9550-15-1-0171 and the National Science Foundation (NSF) under Grants No. DMR-1120923, No. PHY-1415547 and No. ECCS-1158644.
\end{acknowledgments}

\bibliography{ref}

\end{document}